\documentclass[11pt,a4paper]{article}

\pdfoutput=1

\usepackage{booktabs}
\usepackage{dcolumn}  
\usepackage[T1]{fontenc}  
\usepackage[utf8]{inputenc}  
\usepackage{jcappub}
\usepackage{txfonts}  
\usepackage{url}

\newcommand{\lcdm}{$\Lambda$CDM\ }
\newcommand{\hunit}{\text{ km s}^{-1}\text{ Mpc}^{-1}}
\newcommand{\OmegaBo}{\Omega_{\mathrm{b},0}}

\newcommand{\OmegaMo}{\Omega_{\mathrm{m},0}}

\newcommand{\Omegako}{\Omega_{k,0}}
\newcommand{\Omegakdo}{\Omega_{k_\mathrm{d},0}}
\newcommand{\Omegakgo}{\Omega_{k_\mathrm{g},0}}
\newcommand{\be}{\begin{equation}}
\newcommand{\ee}{\end{equation}}
\newcommand{\kg}{{k_\mathrm{g}}}
\newcommand{\kd}{{k_\mathrm{d}}}
\newcommand{\dif}[1]{\mathrm{d}#1}
\newcommand{\abs}[1]{\lvert #1 \rvert}
\newcommand{\emcee}{\texttt{emcee}}
\newcommand{\post}[3]{P(#1 \mid #2, #3)}
\newcommand{\like}[3]{\mathcal{L}(#1 \mid #2, #3)}
\newcommand{\prior}[2]{\mathcal{P}(#1 \mid #2)}
\newcommand{\evid}[2]{\mathcal{E}(#1 \mid #2)}
\newcommand{\N}[2]{\mathcal{N} \left( #1, #2 \right)}
\newcommand{\U}[2]{\mathcal{U} \left( #1, #2 \right)}

\newcolumntype{d}[1]{D{.}{.}{#1}}
\newcolumntype{v}[1]{D{,}{,\ }{#1}}

\title{Testing averaged cosmology with type Ia supernovae and BAO data}
\author[a]{B. Santos,}
\author[b]{A. A. Coley,}
\author[c]{N. Chandrachani Devi}
\author[a]{and J. S. Alcaniz}
\affiliation[a]{
    Departamento de Astronomia, Observatório Nacional, 20921-400, \\
    Rio de Janeiro -- RJ,
    Brasil
}
\affiliation[b]{
    Department of Mathematics and Statistics, Dalhousie University, Halifax, Canada B3H 3J5
}
\affiliation[c]{
    Instituto de Astronomía, Universidad Nacional Autónoma de México, Box 70-264, \\
    México City, México
}
\emailAdd{\tt thoven@on.br}
\emailAdd{\tt aac@mathstat.dal.ca}
\emailAdd{\tt chandrachaniningombam@astro.unam.mx}
\emailAdd{\tt alcaniz@on.br}

\abstract{
    An important problem in precision cosmology is the determination of the effects of averaging and backreaction on observational predictions, particularly in view of the wealth of new observational data and improved statistical techniques.
    In this paper, we discuss the observational viability of a class of averaged cosmologies which consist of a simple parametrized phenomenological two-scale backreaction model with decoupled spatial curvature parameters.
    We perform a Bayesian model selection analysis and find that this class of averaged phenomenological cosmological models is favored with respect to the standard \lcdm cosmological scenario when a joint analysis of current SNe Ia and BAO data is performed.
    In particular, the analysis provides observational evidence for non-trivial spatial curvature.
}


\arxivnumber{1611.01885}

\begin{document}

\maketitle
\flushbottom

\section{\label{sec:introduction}Introduction}

In the standard \lcdm model of cosmology, the Universe is assumed to be described by a single linearly perturbed Friedmann--Lemaître--Robertson--Walker (FLRW) geometry satisfying the Einstein's field equations (EFE) of general relativity (GR).
However, in an inhomogeneous universe, even if the gravitational interaction obeys the EFE on small scales, this will not be true for large-scale averages.
An important problem in cosmology is therefore determining the form of deviations from the EFE when considering geometry averaged on large scales and consequently what effects such backreactions will have on observations~\cite{review}.
Most authors accept that the effects of backreactions will be important for precision cosmology~\cite{NotGW1} (see also~\cite{NotGW2}).
Unfortunately, averaging in GR is a complicated operation, due to the general covariance of the theory and the non-linearity of the EFE.
Indeed, the latter of these ensures that smoothing the spacetime over cosmological scales does not yield the same result as solving the EFE with a smooth matter distribution.

Here we consider a simple phenomenological two-scale cosmological model with a simple parametrized backreaction contribution to the FE, motivated by an exact and fully covariant macroscopic averaging procedure~\cite{Coley2005}.
These models have decoupled spatial curvature parameters in the metric and the Friedmann equation, and the FLRW models of EFE can be easily recovered when these parameters are set to be equal~\cite{CCCS}.
Thus we have a parametrized phenomenological model, similar to other phenomenological extensions of the standard model~\cite{VMS}, within which we can statistically analyze data in order to study the potential non-trivial observational consequences of averaging.

This is the goal of the present paper.
Given our profound lack of understanding of the mechanism behind cosmic acceleration, we study the observational predictions of a class of averaged cosmological models (referred to as the 2CC model hereafter) in the light of the most recent type Ia supernovae (SNe Ia) observations, the so-called Joint Lightcurve Analysis (JLA) sample~\cite{Betoule2014}.
To help break the degeneracy between the model parameters, we also use current measurements of the baryon acoustic oscillations (BAO) scale from a collected sample of BAO measurements taken from various surveys, namely 6dFGS, MGS, BOSS LOWZ, SDSS(R), BOSS CMASS~\cite{Beutler2011, Ross2015, Anderson2014, Padmanabhan2012} and three correlated measurements from the WiggleZ survey~\cite{Blake2012}.
We perform a Bayesian model selection analysis to compare the observational viability of the 2CC model with the standard \lcdm cosmology.
We find that this class of phenomenological cosmological models is favored with respect to the standard scenario when a joint analysis of current SNe Ia and BAO data is performed.

\section{\label{sec:2CC}Two curvature model}

The metric of the spacetime geometry in the phenomenological two curvature cosmological (2CC) model is given as:
\be
    \label{eq:MLE}
    \dif{s}^2 = -\dif{t}^2 + a^2(t) \left[ \frac{\dif{r}^2}{1 - \kg r^2} + r^2 \dif{\Omega} \right] \,,
\ee
with geometrical curvature, $\kg$, and scale factor $a(t)$.
On large scales the macroscopic FE has the form:
\be
    \label{eq:FE}
    H^2 = \frac{\dot{a}^2}{a^2} = \frac{8 \pi G}{3} \rho -\frac{\kd}{a^2} + \frac{\Lambda}{3} \,,
\ee
where the `dynamical curvature', $\kd$, includes contributions from both the spatial curvature and correlations (backreaction).
When the terms that involve backreaction vanish we recover the usual result $\kg = \kd$.
But, in general, in spacetimes that are inhomogeneous on small scales, these two `spatial curvature' terms are not expected to be equal.
After defining the curvature terms as $\Omega_\kg \equiv - \kg / a^2 H^2$ and $\Omega_\kd \equiv - \kd / a^2 H^2$, the macroscopic Friedmann equation then becomes $1 = \Omega_\mathrm{m} + \Omega_\kd + \Omega_\Lambda$, where $\Omega_\mathrm{m}$ and $\Omega_\Lambda$ are the fraction of the energy content of the Universe in matter and the cosmological constant, respectively.
In this model the spatial curvature of the spacetime metric, $\kg$ is decoupled from the spatial curvature $\kd$ that appears in the macroscopic Friedmann equation.
In general, the parameters $\kg$ and $\kd$ could be scale dependent~\cite{CCCS}.

The distance-redshift relation provides the basis for many key observational tests of the cosmological background.
The trajectories of (the average of a large number of) photons are null trajectories with respect to the spacetime metric.
Integrating a null trajectory in the geometry (\ref{eq:MLE}), assuming $\Omega_\kg$ and $\Omega_\kd$ are constant and using the solutions to the macroscopic Friedmann equation (\ref{eq:FE}), gives rise to the luminosity distance-redshift relation:
\be
    \label{eq:dL}
    d_\mathrm{L}(z) = \frac{(1 + z)}{H_0 \sqrt{\abs{\Omegakgo}}}
        f_\kg \left(
            \int_{1/(1+z)}^1 \frac{\sqrt{\abs{\Omegakgo}} \, \dif{a}}{E(a)}
        \right) \,,
\ee
where $E(a) \equiv H(a) / H_0 = \sqrt{\OmegaMo a + \Omegakdo a^2 + \Omega_\Lambda a^4}$ and $f_\kg(x) = \sinh(x)$, $x$ or $\sin(x)$ when $\kg < 0$, $\kg = 0$ or $\kg > 0$, respectively.
This expression reduces to the usual one when $\Omegakgo = \Omegakdo$.

The cosmological probes make observations over a wide range of cosmological scales.
For example, the current observations of BAOs probe the matter distribution at scales of about 150 Mpc, while in the case of high redshift SNe Ia (out to $z \sim 1$), the scales reach out to several Gpc.
On the other hand, the CMB involves making observations on the scale of the horizon ($\sim 14$ Gpc).
The introduction of non-equal, and possibly scale dependent, parameters $\kg$ and $\kd$ will potentially bring us the effects of spatial curvature on the scales of SNe and BAOs, while still satisfying the stringent constraints available on the largest scales from the CMB.
Therefore, a positive detection of conflicting measurements of spatial curvature on different cosmological scales would result to a sign of non-trivial averaging effects that could not be naturally explained by inflation~\cite{COLEYAV} (see also~\cite{DiDio2016, Leonard2016}).

Constraints on the phenomenological two curvature models based on the available data, particularly with the earlier local Hubble rate of $H_0 = 74.2 \pm 3.6 \hunit$, were investigated in~\cite{CCCS}, where the marginalized posterior values of each parameter in the various cases were presented.
It was found that the additional freedom gained by allowing $\Omegakgo \neq \Omegakdo$ is considerable, with constraints on $\Omega_\Lambda$ and the two $\Omega_k$'s being significantly weaker than in the standard approach.
The combination of all of these observables still appears to provide strong evidence for the existence of dark energy.
However, it is striking that constraints on $\Omegakgo$ are an order of magnitude tighter than those on $\Omegakdo$.
There are even tantalizing hints that the data may favor $\Omegakgo \neq \Omegakdo$ (the combination of all data excludes $\Omegakgo = \Omegakdo$ at the 95\% confidence level~\cite{CCCS}).
Presumably better (more optimal) statistical fits are likely utilizing an appropriate framework, not adapted a priori to the standard model.

In view of the plethora of new observational data, especially the recent measurement of the local Hubble constant by Riess et al.~\cite{R16} of $H_0 = 73.00 \pm 1.75 \hunit$ at 68\% c.l., and with improved statistical techniques~\cite{Santos2016}, it is timely to update and revise the constraints on the two curvature models.
Indeed, several authors have recently attempted to explain the significant tension in the Hubble constraint from~\cite{R16} and the Planck data~\cite{planck2015}; for example, a combined analysis of all of the recent data in an extended 12 cosmological parameter space was presented in~\cite{VMS} (see also~\cite{Racz} and~\cite{moffat}).

\section{\label{sec:data}Data sets}

In this analysis, we use two different data sets to test the observational viability of the 2CC cosmology.
We describe them in what follows.

\subsection{\label{subsec:sneia}Type Ia supernovae}

Observations of type Ia supernovae (SNe Ia) are a key probe of the cosmic expansion on large-scales, in that they map the expansion history of the Universe up to $z \lesssim 2$.
The most attractive feature of these events is that their absolute magnitude can be approximated by using light-curve templates to extract their `stretch' and `color' parameters, enabling them to be considered as ``standardizable candles''.
We use the Joint Light-Curve Analysis (JLA)~\cite{Betoule2014} sample which is an extension of the compilation provided by ref.~\cite{Conley2011}.
It contains a set of 740 spectroscopically confirmed SNe Ia composed by several low-redshift ($z < 0.1$) samples, the full three-year SDSS-II supernova survey~\cite{Sako2014} sample with redshift $0.05 < z < 0.4$, the three-years data of the SNLS survey~\cite{Conley2011, Guy2010} up to redshift $z < 1$ and a few high redshift Hubble Space Telescope (HST) SNe~\cite{Riess2007} in the interval $0.216 < z < 1.755$.
The photometry of SDSS and SNLS was re-calibrated and the SALT2 model is retrained using the joint data set.

From the observational point of view, the distance modulus of a SN Ia is obtained by a linear relation from its light-curve:
\be
    \label{eq:distmod_obs}
    \mu = m_B - (M_B - \alpha \times x_1 + \beta \times c) \,,
\ee
where $m_B$ is the observed B-band peak magnitude, $x_1$ is the time stretching of the light-curve, and $c$ is the supernova color at maximum brightness.
These three light-curve parameters, $m_B$, $x_1$ and $c$, have different values for each supernova and are derived directly from the light-curves.
The nuisance parameters $\alpha$, $\beta$ and $M_B$ describe the shape and color corrections of the light-curve, and the absolute magnitude of the SN Ia, respectively.
They are assumed to be constants for all the supernovae, but different for different cosmological models.
To model the effect of host galaxy properties on the SN Ia intrinsic brightness, we follow ref.~\cite{Betoule2014} and assume a step function relation between $M_B$ and the host galaxy stellar mass, $M_\text{host}$, in that $M_B \rightarrow M_B + \Delta_M$ for $\log_{10} M_\text{host} > 10$.
Thus, the nuisance parameters corresponding to the JLA measurements are $\alpha$, $\beta$, $M_B$ and $\Delta_M$.

\subsection{\label{subsec:bao}Baryon acoustic oscillations}

Another key tool to probe the expansion rate and the large-scale properties of the Universe is the observation of baryon acoustic oscillations (BAO), which are the imprints in the large-scale structure of matter due to the oscillations in the primordial plasma.
The BAO measurements and their calibration with CMB anisotropy data provide a powerful standard ruler to probe the angular-diameter distance versus redshift relation and the Hubble parameter evolution.

Table~\ref{tab:data_BAO} shows the BAO distance measurements employed in this work.
The distance-redshift relation related to BAO measurements is usually obtained by performing a spherical average of the BAO scale measurement and is given by
\be
    \label{eq:dz}
    d_z = \frac{r_s(z_\text{drag})}{D_V(z)} \,,
\ee
where
\be
    \label{eq:rs}
    r_s(z_\text{drag}) = \frac{c}{\sqrt{3}} \int_{z_\text{drag}}^\infty \frac{\dif{z}}{\sqrt{1 + (3\OmegaBo / 4\Omega_{\gamma,0}) (1 + z)^{-1}} H(z)}
\ee
is the radius of the comoving sound horizon at the drag epoch $z_\text{drag}$ when photons and baryons decouple~\cite{Eisenstein1998}, $D_V(z) = \left[ cz d_C^2(z) / H(z) \right]^{1/3}$ is the volume-averaged distance~\cite{Eisenstein2005} and $d_C(z) = d_\mathrm{L}(z) / (1 + z)$, with $d_\mathrm{L}(z)$ given by eq.~(\ref{eq:dL}), is the comoving angular diameter distance.
In~(\ref{eq:rs}), $\OmegaBo$ and $\Omega_{\gamma,0}$ are the present values of the baryon and photon density parameters, respectively.
In this work we use $\OmegaBo = 0.022765 h^{-2}$ and $\Omega_{\gamma,0} = 2.469 \times 10^{-5} h^{-2}$ as given by ref.~\cite{WMAP2011}, where $h \equiv H_0 / 100 \hunit$.

In the table~\ref{tab:data_BAO}, the three measurements obtained from the WiggleZ survey are correlated.
So, when using those data points, one must include their inverse covariance matrix, which is:
\be
    \label{eq:WiggleZ_cov}
    C^{-1} =
    \begin{pmatrix}
        1040.3 & -807.5  & 336.8 \\
        -807.5 & 3720.3  & -1551.9 \\
        336.8  & -1551.9 & 2914.9
    \end{pmatrix} \,.
\ee

\begin{table}[tbp]
    \centering
    \begin{tabular}{|c|ccc|}
        \hline
        Survey     & $z$     & $d_z(z)$            & Reference \\
        \hline
        6dFGS      & $0.106$ & $0.3360 \pm 0.0150$ & \cite{Beutler2011} \\
        MGS        & $0.15$  & $0.2239 \pm 0.0084$ & \cite{Ross2015} \\
        BOSS LOWZ  & $0.32$  & $0.1181 \pm 0.0024$ & \cite{Anderson2014} \\
        SDSS(R)    & $0.35$  & $0.1126 \pm 0.0022$ & \cite{Padmanabhan2012} \\
        BOSS CMASS & $0.57$  & $0.0726 \pm 0.0007$ & \cite{Anderson2014} \\
        WiggleZ    & $0.44$  & $0.073$             & \cite{Blake2012} \\
        WiggleZ    & $0.6$   & $0.0726$            & \cite{Blake2012} \\
        WiggleZ    & $0.73$  & $0.0592$            & \cite{Blake2012} \\
        \hline
    \end{tabular}
    \caption{\label{tab:data_BAO}BAO measurements used in this work.}
\end{table}

\section{\label{sec:method}Method}

In what follows, we apply the Bayesian framework to perform a parameter estimation for the 2CC model and a model comparison between the $\kd \neq \kg$ (2CC) and $\kd = \kg$ ($\Lambda$CDM) scenarios.

\subsection{\label{subsec:estimation}Parameter estimation}

In Bayesian inference, all data analysis for a given dataset $D$ is performed by computing the joint posterior for a set $\Theta$ of free parameters through the Bayes' Theorem~\cite{Bayes1764}:
\be
    \label{eq:BayesTheorem}
    \post{\Theta}{D}{M} = \dfrac{\like{D}{\Theta}{M} \, \prior{\Theta}{M}}{\evid{D}{M}} \,,
\ee
where $P$, $\mathcal{L}$, $\mathcal{P}$ and $\mathcal{E}$ are the shorthands for the \emph{posterior}, the \emph{likelihood}, the \emph{prior} and the \emph{Evidence}, respectively.
Assuming we have a set of physically interesting parameters $\theta$ and a set of nuisance parameters $\phi$, and that the full set of parameters is $\Theta = (\theta, \phi)$, we can write the posterior~(\ref{eq:BayesTheorem}) on the parameter of interest marginalized over the nuisance parameters as
\be
    \label{eq:BT_parameter}
    \post{\theta}{D}{M} \propto \int \like{D}{\Theta}{M} \, \prior{\Theta}{M} \, \dif{\phi} \,,
\ee
where the proportionality symbol ``$\propto$'' is due to the fact that the Evidence in~(\ref{eq:BayesTheorem}) is a normalization constant and thus irrelevant in parameter estimation.

For the JLA SNe Ia sample, we assume a multivariate Gaussian likelihood of the type
\be
    \mathcal{L}_\text{JLA}(D \mid \Theta) = \exp[-\chi_\text{JLA}^2(D \mid \Theta) / 2] \,,
\ee
with
\be
    \label{eq:chi2JLA}
    \chi^2_\text{JLA}(D \mid \Theta) = \left[ \mathbf{m}_B - \mathbf{m}_B(\Theta) \right]^T C^{-1} \left[ \mathbf{m}_B - \mathbf{m}_B(\Theta) \right] \,,
\ee
where $\mathbf{m}_B$ is the vector of observed B-band magnitude measurements (table F.3 of ref.~\cite{Betoule2014}), $\mathbf{m}_B(\Theta)$ is the vector with the predicted ones which can be obtained from Eqs.~(\ref{eq:dL}) and~(\ref{eq:distmod_obs}) by using the theoretical distance modulus, $\mu(z ; \Theta) = 5\log{\left[ d_\mathrm{L}(z ; \Theta) / 10\,\mathrm{pc} \right]}$.
The matrix $C$ corresponds to the covariance matrix of the distance modulus $\mu$, estimated accounting various statistical and systematic uncertainties (we refer the reader to ref.~\cite{Betoule2014} for more information about these uncertainties).

Using the same methodology applied to the JLA compilation, we also consider a multivariate Gaussian likelihood for the BAO data set.
For each survey listed in the first column of the table~\ref{tab:data_BAO}, except the WiggleZ survey, the chi-square is given by
\be
    \label{eq:chi2BAO_table}
    \chi_\text{survey}^2(D \mid \Theta) = \left[ \dfrac{d_{z,\text{survey}} - d_z(z_\text{survey}; \Theta)}{\sigma_\text{survey}} \right]^2 \,,
\ee
where $d_{z,\text{survey}}$ and $d_z(z_\text{survey}; \Theta)$ are the observed and theoretical $d_z$, respectively, and $\sigma_\text{survey}$ is the uncertainty associated with each observed value.
For the WiggleZ data we use the matrix $C$ given by eq.~(\ref{eq:WiggleZ_cov}) and write the chi-square in the form
\be
    \label{eq:chi2BAO_WiggleZ}
    \chi_\text{WiggleZ}^2(D \mid \Theta) = \left[ \mathbf{d}_{z,i} - \mathbf{d}_z(\Theta) \right]^T C^{-1} \left[ \mathbf{d}_{z,i} - \mathbf{d}_z(\Theta) \right] \,.
\ee
Then, the BAO likelihood is directly obtained by the product of the individual likelihoods as $\mathcal{L}_\text{BAO} = \mathcal{L}_\text{6dFGS} \times \mathcal{L}_\text{MGS} \times \mathcal{L}_\text{LOWZ} \times \mathcal{L}_\text{SDSS(R)} \times \mathcal{L}_\text{CMASS} \times \mathcal{L}_\text{WiggleZ}$.
Similarly, the joint likelihood for the JLA SNe Ia compilation and the BAO data is given by $\mathcal{L}_\text{joint} = \mathcal{L}_\text{JLA} \times \mathcal{L}_\text{BAO}$.

It can be seen from~(\ref{eq:BayesTheorem}) that the Bayes' Theorem updates our previous knowledge about some model parameters in the light of a given data set.
Hence, the dependence on the priors $\prior{\Theta}{M}$ chosen for the free parameters is an intrinsic feature of Bayesian inference (both parameter estimation and model selection) and accounts for the model's predictive power.
We apply the results shown in ref.~\cite{CCCS} for the CMB + HST + Union2 + BAO analysis as the Gaussian prior probabilities for $\OmegaMo$, $\Omegakdo$ and $\Omegakgo$.
We also assume a Gaussian prior for the Hubble constant of $H_0 = 73.24 \pm 1.74 \hunit$, a 2.4\% determination given recently by ref.~\cite{R16}.
However, for the analysis involving the JLA dataset, we applied uniform priors for $\alpha$, $\beta$, $M_B$ and $\Delta_M$, considered in this work as nuisance parameters.
All priors used are shown in the table~\ref{tab:priors}.

\begin{table}[tbp]
    \centering
    \begin{tabular}{|c|l|}
        \hline
        Parameter   & Prior \\
        \hline
        $\OmegaMo$  & $\N{0.282}{0.01}$ \\
        $\Omegakdo$ & $\N{-0.033}{0.005}$ \\
        $\Omegakgo$ & $\N{-0.004}{0.0001}$ \\
        $H_0$       & $\N{73.24}{3.028}$ \\
        $\alpha$    & $\U{0}{0.5}$ \\
        $\beta$     & $\U{0}{5}$ \\
        $M_B$       & $\U{-30}{-10}$ \\
        $\Delta_M$  & $\U{-0.5}{0.5}$ \\
        \hline
    \end{tabular}
    \caption{\label{tab:priors}Priors on the parameters of the 2CC model, where $\N{\mu}{\sigma^2}$ and $\U{a}{b}$ are the Gaussian (with mean $\mu$ and variance $\sigma^2$) and the uniform priors, respectively. Note that, since $\U{a}{b}$ is normalized to unity, for this kind of prior we have $\mathcal{P}(x | M) = 1 / (b - a)$ for $a \leq x \leq b$ and $\mathcal{P}(x | M) = 0$ otherwise.}
\end{table}

Due to the difficulty in computing the posterior in~(\ref{eq:BayesTheorem}) both analytically and numerically, Markov Chain Monte Carlo (MCMC) sampling techniques are nowadays widely applied for this task (we refer the reader to refs.~\cite{Metropolis1953, Mackay2003, Skilling2004, Feroz2013} for some MCMC algorithms and to refs.~\cite{Lewis2002, Mukherjee2006} for applications of some of those algorithms in cosmology).
In this work, we used the Affine-Invariant MCMC Ensemble sampler~\cite{Goodman2010} through the Python package {\emcee}~\cite{ForemanMackey2013}.
The name of this sampler is due to its performance invariance under linear transformations of the parameter space, which makes it to be a good tool to sample from a large number of different kinds of distributions.

This method works by moving several chains (or \emph{walkers}) in parallel through the parameter space.
For all of the analysis, we chose to work with 250 walkers for 400 iterations (steps) to get a sample with $N = 10^5$ points.
To ensure that initialisation was forgotten, before these 400 iterations, we also ran a burn-in phase analysing the exponential autocorrelation time $\tau_\mathrm{exp}$ at each iteration.
During this phase, we applied a very conservative stopping criterion~\cite{Allison2014} in that the burn-in phase stops if $i > 20 \times \max(\tau_\mathrm{exp})$, where $i$ is the iteration at the burn-in phase and $\max(\tau_{\mathrm{exp},i})$ is the maximum $\tau_{\mathrm{exp},i}$ among all the dimensions of the parameter space at the i-\emph{th} burn-in iteration.
However, we did not set any stopping criterion for the 400 iterations after the burn-in phase, but we verified that the statistical error $\varepsilon = \sqrt{2\tau_\mathrm{int} / N}$\footnote{$\tau_\mathrm{int}$ is the integrated autocorrelation time (see ref.~\cite{Sokal1997} for more information).} of any parameter was, in all cases, lower than 2.4\% of the standard deviation of its marginal distribution, indicating a good convergence of all samples.

\subsection{\label{subsec:lcdm_comp}Comparison to \lcdm}

Although being uninteresting for parameter estimation, the Evidence $\mathcal{E}$ in~(\ref{eq:BayesTheorem}) is crucial for model comparison.
It evaluates the model's performance in the light of the data by integrating the product $\mathcal{L}\,\mathcal{P}$ over the full parametric space of the model:
\be
    \label{eq:evidence}
    \evid{D}{M} = \int_M \like{D}{\Theta}{M} \, \prior{\Theta}{M} \, \dif{\Theta} \,.
\ee
Therefore, the Evidence gives the probability of obtaining the data $D$ in the context of a given model $M$.

Since in many cases it is very difficult to integrate~(\ref{eq:evidence}) numerically, the {\emcee} package computes the Evidence using the Parallel-Tempered MCMC~\cite{Swendsen1986, Earl2005}, which works by sampling from a modified posterior $P \propto \mathcal{L}^{\beta} \mathcal{P}$, running multiple chains at different ``temperatures'' $T = 1 / \beta > 1$.
Only the chain for which $\beta = 1$ is used for inference, since this corresponds to the case where the usual posterior of the Bayes' Theorem~(\ref{eq:BayesTheorem}) is recovered.
The Evidence is estimated by this method after computing the average of the modified log-likelihood at the different temperatures and applying the Thermodynamic Integration technique\footnote{More information about how {\emcee} estimates the Evidence can be found at \url{http://dan.iel.fm/emcee/current/user/pt}.}~\cite{Goggans2004} to approximate the following integral~\cite{ForemanMackey2013}:
\be
    \ln{\mathcal{E}(\beta=1)} = \int_0^1 \langle \ln{\mathcal{L}} \rangle_\beta \, \dif{\beta}. \,
\ee
Thus, this method transforms the multiple integral in~(\ref{eq:evidence}) into a one-dimensional integral, at the cost of increasing the number of chains.
At this point, it is important to comment that the number of temperatures impacts directly the numerical uncertainty in the Evidence calculation.
A very small number of temperatures might lead to a poor estimate of the integral, while a larger number of temperatures will yield smaller uncertainty, but will proportionally increase the computation time.
We perform all analysis using 10 temperatures as we have verified that this is a good number for the purpose of this work.
As an example, changing the number of temperatures from 10 to 20 (the {\emcee} default value) leads to an execution time $2\times$ slower and improves the value of $\ln{\mathcal{E}}$ by no better than 0.06\% (and no more than 0.6\% in its uncertainty).

To discriminate between the 2CC model and the standard \lcdm scenario, we computed the Bayes Factor of the \lcdm model relative to the 2CC model, given by $B \equiv \mathcal{E}_{\Lambda\text{CDM}} / \mathcal{E}_\mathrm{2CC}$, and adopted the following scale to interpret the values of $\ln{B}$: values of $\abs{\ln{B}}< 1$ indicate an \emph{inconclusive} evidence whereas values of ${\ln{B}}$ above 1, 2.5 and 5 indicate a \emph{weak}, \emph{moderate} and \emph{strong} evidence in favor of the \lcdm model, respectively.
This is a conservative version of the so-called Jeffreys' Scale~\cite{Jeffreys1961}, as suggested by ref.~\cite{Trotta2008}.
Note that $\ln{B} < -1$ means support in favor of the 2CC model.

Finally, regarding the choice of the priors for the parameters of the \lcdm model, we followed the same methodology applied to obtain the priors for the 2CC parameters, i.e., using the results presented by ref.~\cite{CCCS} but now for the $\kd = \kg = k$ case.
Therefore, we assumed the Gaussian priors $\OmegaMo = 0.277 \pm 0.017$ and $\Omegako = 0.0 \pm 0.006$ for the matter and curvature density parameters, respectively.
The priors on $H_0$ and on the JLA nuisance parameters were kept the same as those shown in table~\ref{tab:priors}.

\section{\label{sec:results}Results}

\begin{figure}[tbp]
    \centering
    \includegraphics[width=\columnwidth]{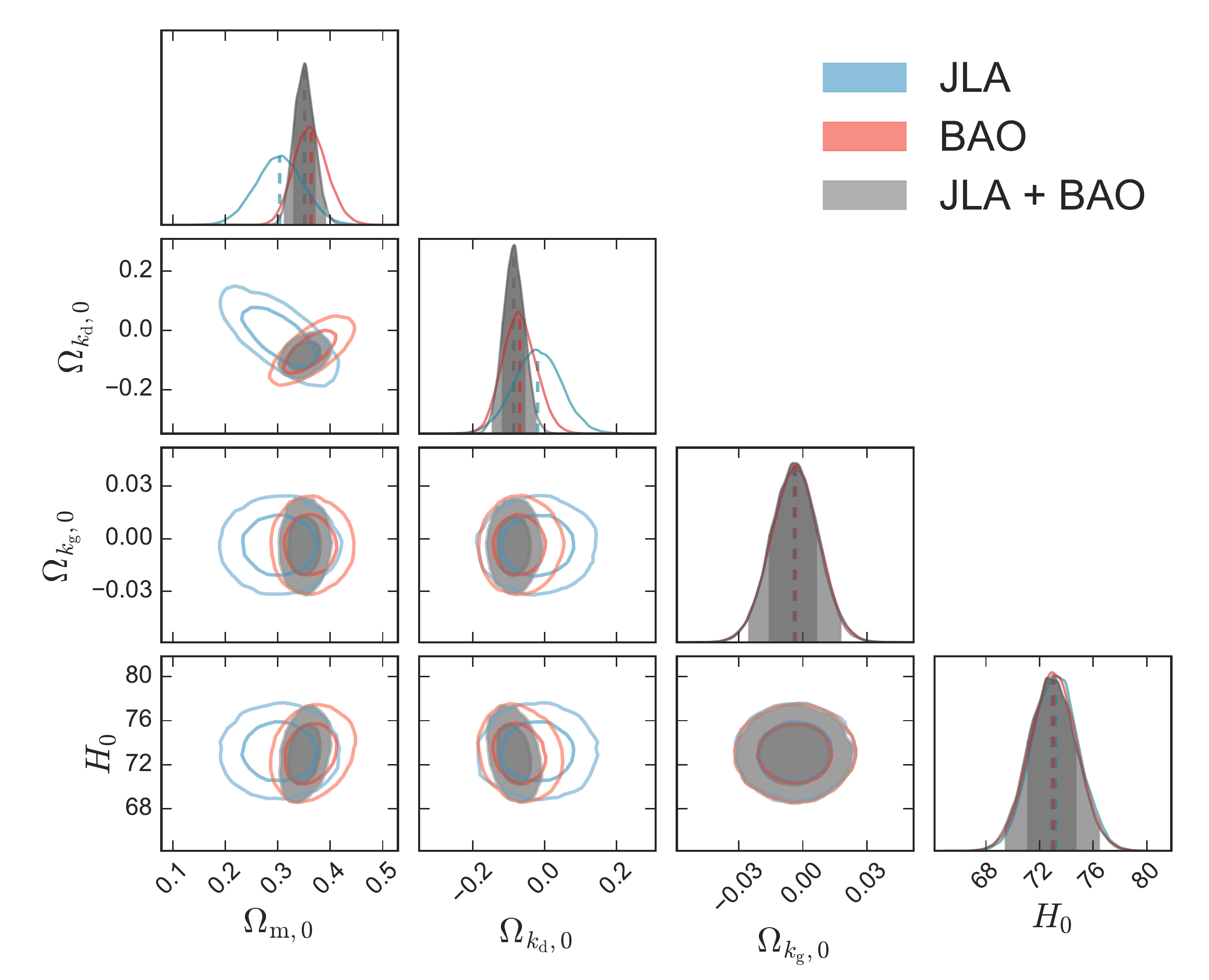}
    \caption{\label{fig:contours}Marginalized posterior distributions and credible intervals (68\% and 95\%) for the 2CC model.}
\end{figure}

Table~\ref{tab:results_estimates} shows individual and combined constraints on the parameters of the 2CC model, with the first, second and third sub-tables corresponding to the results obtained from SNe Ia, BAO and SNe Ia + BAO data, respectively.
We note that regardless of the data set used in the analysis, the constraints on $\Omegakgo$ are an order of magnitude tighter than those on $\Omegakdo$, which is in good agreement with previous results~\cite{CCCS}.
Clearly, the combination of data favors values of $\Omegakgo \neq \Omegakdo$, with the standard case, $\Omegakgo = \Omegakdo$, being off by $\sim 95\%$ credible interval.
For completeness, we also show the credible intervals (68\% and 95\%) for all combinations of the 2CC parameters in the figure~\ref{fig:contours}.

\begin{table}[tbp]
    \centering
    \begin{tabular}{| c | d{3.3} d{3.3} v{8.9} |}
        \hline
        Parameter & \multicolumn{1}{c}{Mean} & \multicolumn{1}{c}{Std. dev.} & \multicolumn{1}{c|}{95\% c.i.} \\
        \hline
        \multicolumn{4}{|c|}{JLA} \\
        \hline
        $\OmegaMo$  & 0.304  & 0.045 & (0.212, 0.393) \\
        $\Omegakdo$ & -0.020 & 0.066 & (-0.150, 0.113) \\
        $\Omegakgo$ & -0.004 & 0.011 & (-0.026, 0.018) \\
        $H_0$       & 73.188 & 1.730 & (69.709, 76.581) \\
        \hline
        \multicolumn{4}{|c|}{BAO} \\
        \hline
        $\OmegaMo$  & 0.363  & 0.032 & (0.300, 0.427) \\
        $\Omegakdo$ & -0.070 & 0.046 & (-0.160, 0.024) \\
        $\Omegakgo$ & -0.004 & 0.011 & (-0.026, 0.018) \\
        $H_0$       & 73.053 & 1.738 & (69.608, 76.523) \\
        \hline
        \multicolumn{4}{|c|}{JLA + BAO} \\
        \hline
        $\OmegaMo$  & 0.351  & 0.020 & (0.313, 0.391) \\
        $\Omegakdo$ & -0.085 & 0.030 & (-0.145, -0.026) \\
        $\Omegakgo$ & -0.004 & 0.011 & (-0.025, 0.018) \\
        $H_0$       & 72.946 & 1.752 & (69.477, 76.427) \\
        \hline
    \end{tabular}
    \caption{\label{tab:results_estimates}Estimates of the parameters of the 2CC model.
    The last column shows the 95\% High Posterior Density credible intervals.}
\end{table}

\begin{table}[tbp]
    \centering
    \begin{tabular}{| c | d{4.10} c |}
        \hline
        Model & \multicolumn{1}{c}{$\ln{\mathcal{E}}$} & \multicolumn{1}{c|}{$\ln{B}$} \\
        \hline
        \multicolumn{3}{|c|}{JLA} \\
        \hline
        2CC   & -352.247 \pm 2.784 & $0.884 \pm 3.865$ \\
        \lcdm & -351.363 \pm 2.681 & $0$ \\
        \hline
        \multicolumn{3}{|c|}{BAO} \\
        \hline
        2CC   & -4.665 \pm 0.435   & $-5.810 \pm 1.912$ \\
        \lcdm & -10.476 \pm 1.862  & $0$ \\
        \hline
        \multicolumn{3}{|c|}{JLA + BAO} \\
        \hline
        2CC   & -356.810 \pm 3.124 & $-6.549 \pm 5.264$ \\
        \lcdm & -363.359 \pm 4.237 & $0$ \\
        \hline
    \end{tabular}
    \caption{\label{tab:results_modelselection}Bayesian evidence and Bayes factors for the \lcdm model related to the 2CC model.}
\end{table}

\begin{figure}[tbp]
    \centering
    \includegraphics[width=0.8\columnwidth]{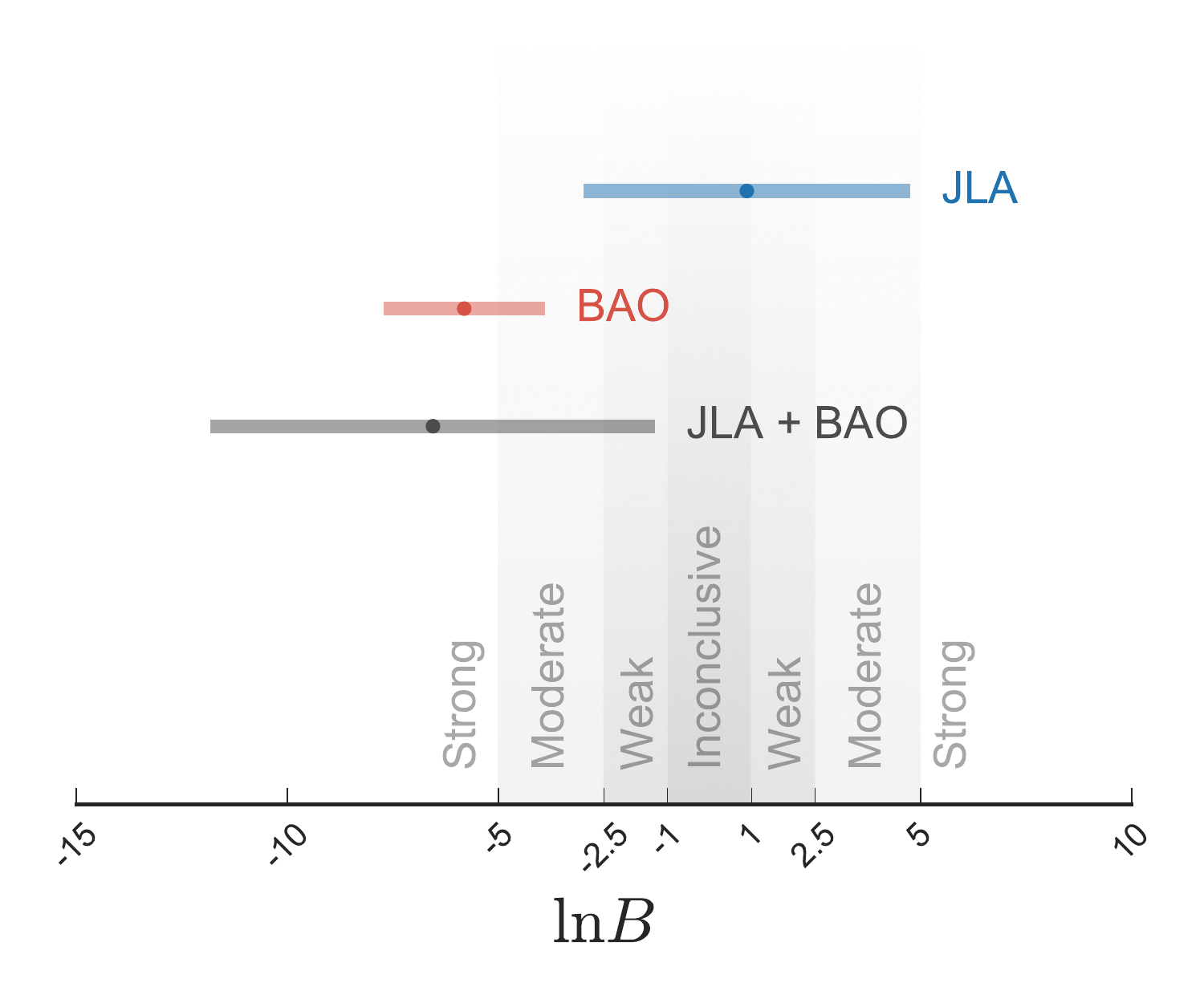}
    \caption{\label{fig:logBFs}Intervals for the Bayes factors between \lcdm and the 2CC model given in table~\ref{tab:results_modelselection}. Note that the standard \lcdm scenario is favored when $\ln{B} > 1$.}
\end{figure}

The main quantitative results of our Bayesian model selection analysis are shown in table~\ref{tab:results_modelselection}.
From these results, we observe that the current SNe Ia data alone cannot distinguish between the 2CC and standard models, with $\ln{B} = 0.884 \pm 3.865$, which varies from moderate evidence against to moderate evidence in favor of the \lcdm scenario.
On the other hand, the BAO and the SNe Ia + BAO data are much more effective to this end, as can be seen in the second and third sub-tables of table~\ref{tab:results_modelselection}.
For the joint analysis of SNe Ia and BAO measurements, we find $\ln{B} = -6.549 \pm 5.264$, which indicates that the evidence of the standard \lcdm model varies from strongly to moderately disfavored with respect to the 2CC cosmology.
A graphical representation of the ranges of all Bayes factors is displayed in the figure~\ref{fig:logBFs}.
It is worth mentioning that we also explored the dependence of the results on a different prior on $H_0$~\cite{planck2015} and we find that the results of table~\ref{tab:results_modelselection} remain unchanged.

These results can be compared with those presented in ref.~\cite{Santos2016}, in which a Bayesian comparison of different classes of alternative cosmologies was performed.
Using eq.~(\ref{eq:evidence}) and applying the same uniform priors of that reference on the nuisance parameters $\alpha$, $\beta$, $M_B$ and $\Delta_M$, the Bayesian evidence in the last sub-table of table~\ref{tab:results_modelselection} changes to $\ln{E} = -351.570 \pm 3.124$ and $\ln{E} = -358.119 \pm 4.237$ for the 2CC and \lcdm models, respectively.
Keeping in mind that the Gaussian prior on the $\OmegaMo$ parameter is different from the prior used in the ref.~\cite{Santos2016} and that the \lcdm model investigated in this work is non-flat ($k \neq 0$), these results would put the 2CC model on the top of the rank shown in the last sub-table of table 5 of ref.~\cite{Santos2016}.
Moreover, we do not expect that the values of $\ln{E}$ for the 2CC model would change significantly if we also had corrected the Gaussian priors, as can be seen from the $\sim 2.2\%$ difference between the $\ln{E}$ value for \lcdm using same uniform priors of ref.~\cite{Santos2016} (-358.119) and the value shown in the last sub-table of table 5 of the same reference.

\section{\label{sec:conclusions}Conclusions}

A critical problem in cosmology consists of determining the exact form of deviations from the EFE when considering geometry averaged on large scales and their effects on observational predictions.
While a consensus is lacking on this latter aspect, some results suggest that although not responsible for cosmic acceleration, averaging may be important for precision cosmology~\cite{Coley2005, Kolb2005}.

In this paper we have discussed the observational viability of a class of averaged cosmologies which consists of a simple phenomenological two-scale model with a simple parametrized backreaction contribution to the EFE, motivated by an exact and fully covariant macroscopic averaging procedure~\cite{Coley2005}.
Using the most recent SNe Ia data and current measurements of the BAO scale, we have discussed the effect of allowing the curvature parameters $\Omegakdo$ and $\Omegakgo$ to be independent and shown its consequences for parameter estimation.
We have found that the combination of these data sets requires values of $\Omegakdo \neq \Omegakgo$, with the constraints on the spatial curvature parameter appearing in the Friedmann equation ($\Omegakdo$) being significantly weaker than those on $\Omegakgo$, which appears in the macroscopic metric.

We have also performed a Bayesian model selection statistics to compare the predictivity power of the averaged model with respect to the standard \lcdm cosmology.
From this analysis, we have found that although the current SNe Ia data alone cannot distinguish between these two models ($\ln{B} = 0.884 \pm 3.865$) the combination of SNe Ia and BAO data sets provides $\ln{B} = -6.549 \pm 5.264$, which favors significantly the 2CC model with respect to the standard cosmology.

We do not necessarily present these results as evidence in favor of the simple phenomenological model discussed here, but rather more as motivation for studying more physical models that take into consideration the possible effects of backreaction which can affect cosmological observations at the level of a few percent.
In particular, the analysis suggests observational evidence for non-trivial spatial curvature above the cosmic variance limit (beyond which constraints cannot be meaningfully improved due to the cosmic variance of horizon-scale perturbations), which could be the result of general relativistic effects on the large-scale structure.
These results are timely, since future precision measurements of the spatial curvature are affected by relativistic effects (which are not normally taken into account in standard analyses and which may consequently lead to strongly biased constraints on the curvature parameter)~\cite{DiDio2016}, and the assumption of flatness would preclude a number of potentially powerful tests of early Universe physics~\cite{Leonard2016}.

Finally, it is also worth mentioning that although the two datasets applied in this work show observational evidence for non-trivial spatial curvature, a more robust analysis could be performed by the inclusion of other cosmological observables.
Indeed, we are currently engaged in performing a Bayesian study using current CMB and LSS observations.
Since this kind of analysis requires a non-trivial set of perturbation equations for the 2CC model, we intend to present the results in a forthcoming communication.

\acknowledgments
    A. Coley acknowledges the support of NSERC.
    B. Santos and N. Chandrachani Devi are supported by the National Observatory DTI-PCI program of the Brazilian Ministry of Science, Technology and Innovation (MCTI) and by the DGAPA-UNAM post-doctoral fellowship, respectively.
    J. S. Alcaniz acknowledges CNPq and FAPERJ for financial support.

\bibliographystyle{JHEP}
\bibliography{ref}

\end{document}